\documentclass[journal]{IEEEtran}
\ifCLASSINFOpdf
\else
\fi
\usepackage{color}
\usepackage{graphicx}   
\begin{document}
%
\title{Investigation of deep levels in CdZnTeSe  crystal and their effect on the internal electric field of CdZnTeSe gamma-ray detector}
%
%
%

\author{M Rejhon, V D\v{e}di\v{c}, L Beran, U N Roy, J Franc and R B James
\thanks{M Rejhon,V D\v{e}di\v{c}, L Beran and J Franc are with Charles University, Faculty of Mathematics and Physics, Institute of Physics, Ke Karlovu 5, CZ-121 16 Prague 2, Czech Republic, e-mail: rejhonm@karlov.mff.cuni.cz}
\thanks{U N Roy is with Nonproliferation and National Security Department, Brookhaven National Laboratory, Upton, New York, USA.}
\thanks{R B James is with Savannah River National Laboratory, Aiken, South Carolina 29808, USA.}
\thanks{Manuscript received August, 2018.}}

\maketitle

\begin{abstract}
\textcolor{black}{We present a study of deep levels in CdZnTeSe radiation-detection materials. The approach relies on electrical methods that combine time and temperature evolution of the electric field and electric current after switching on the bias voltage.} We also applied two optical methods to study the deep levels. The first method utilizes the temperature and temporal analysis of the electric-field evolution after switching off an additional light illuminating the sample at a wavelength of 940 nm. \textcolor{black}{The second method involved measuring of the electric-field spectral dependence during near infrared illumination.} We compare the results with those obtained with high-quality CdZnTe detector-grade material. We conclude that the introduction of Se into the lattice leads to a shift of the second ionization level of the Cd vacancy towards the conduction band, as predicted \textcolor{black}{recently by first-principles calculations based on screened hybrid functionals}.
\end{abstract}


%
\IEEEpeerreviewmaketitle

\section{Introduction}
%
%
%
%
\IEEEPARstart{T}{he} CdTe and CdZnTe materials have been effectively used for the production of X-ray and gamma-ray
detectors operating at room temperature for applications in medical imaging, security and astrophysics \cite{SCHLESINGER2001103,Iniewski2016,WAHL2015377,Krawczynski2016}. However, there are still several drawbacks that limit the yield of detector-grade material in the crystals. These include mainly the presence of sub-grain boundaries, Te inclusions/precipitates and compositional inhomogeneity arising from the non-unity segregation coefficient of Zn in the CdTe matrix \cite{SCHLESINGER2001103,Szeles2002,Bolotnikov2010,MacKenzie2013,BURGER2000586}. 

The sub-grain boundaries are formed during the post-growth ingot cooling process due to the poor thermo-physical properties of the melt and the solidified material. These boundaries act as trapping centres and affect the electrical and transport properties of the material \cite{BUIS2014188, BOLOTNIKOV201346}.

The Te inclusions/precipitates originate from the growth process under a Te-rich atmosphere. These secondary phases affect the quality and performance of the CdTe and CdZnTe (CZT) depending on their concentration and size \cite{Bolotnikov2010}. Also, Te secondary phases act as trapping and recombination centres and are responsible for a degradation of the material's detection parameters \cite{CARINI2007120,Gul2016}.

Recently it has been shown that the addition of Se in the matrix results in an effective lattice hardening and a decreased Te inclusion/precipitate concentration with increased Se content. It was observed that CdZnTeSe (CZTS) crystals exhibit better crystallinity than CZT, which can lead to a larger yield of high-quality material with comparable electrical and spectroscopic properties as for CdTe and CZT materials \cite{Gul2017}.

In the current study, we seek to determine the deep levels that participate in space-charge formation and determine the internal electric-field  profile and its dynamics after application of a bias and incident radiation flux. We compare the results with our previous study of high-quality CdZnTe detector material to access the impact of Se on the defect structure. We applied two optical methods to study the deep levels. The first method was the temperature and temporal analysis of the electric-field evolution after bias application and after switching off an additional light illuminating the sample at a wavelength at $940$~nm. The light source with this photon energy forms a positive space charge in the sample, which causes the optically induced polarization. The second method was the scanning of the electric-field response to incident infrared (NIR) illumination in the range of $900-1800$~nm. 

 

\section{Experimental}
For this study we have chosen a CZTS sample with $10\%$ Zn and $4\%$ Se content. The material was doped with In and grown by the travelling heater method. The sample dimensions are $6.50\times5.30\times2.68$~mm$^{3}$. The sample surfaces were mechanically polished with Al$_{2}$O$_{3}$ abrasive (surface RMS 2~nm) without any further chemical treatment. Gold and indium contacts were deposited on the large opposite sides by evaporation. \textcolor{black}{ The gold contact was equipped with a guard ring to separate the effects of the bulk and surface leakage currents during electrical measurements}.

\textcolor{black}{The sample's energy band-gap was evaluated by a spectroscopic ellipsometry measurement using a commercial JA Woollam Co. RC2 ellipsometer. The ellipsometry was measured in reflection mode with three incident angles of $55^{\circ}$ , $60^{\circ}$ and $65^{\circ}$,
respectively. The range of spectral photon energies was between $0.80$ and $6.42$~eV.}

The dependence of a steady-state electric field on the photon energy for NIR illumination was measured using the Pockels effect. A standard setup consisted of a test light, two polarizers and an InGaAs camera. The sample was placed between two crossed polarizers, and the test light operating at a wavelength of $1550$~nm was passed through the sample. The transmitted light $I(x,y)$ was detected by an InGaAs camera. The distribution of the internal electric field is $E(x,y)\approx\sqrt{I(x,y)}$. More details about the measurement setup are provided in our previous articles \cite{rejhonJPD,Franc2015}.

The light from the monochromator in the range $900-1800$~nm was used to scan the spectral dependence of the internal electric field in the sample with a step size of $25$~nm. The monochromatic incident light had a constant photon flux of $1.0\times10^{15}$ photons cm$^{-2}$s$^{-1}$, and the measurements were carried out at room temperature (300 K). This technique allows one to determine the deep levels in the detector due to the measurement of the change of the internal space charge \cite{Franc2015,Zazvorka2014}.

We measured the temporal and temperature evolutions of the electric field after application of the bias and switching off the LED at $940$~nm and a photon flux of $5.5\times10^{15}$ photons cm$^{-2}$s$^{-1}$. The electric field was measured using the same standard setup as for the previous experiment. The incident light at the wavelength of $940$~nm forms a positive space charge in the CdTe and CZT detectors, which induced polarization of the detectors \cite{Franc2015, Dedic2015, Dedic2012}. The return of the electric field to the original state (i.e., without illumination) based on the thermal emission of charge carriers from the traps allows determination of the deep levels responsible for the polarization \cite{rejhonJPD, Dedic2017}. \textcolor{black}{We also employed an analysis of the time and temperature evolution of the electric current after switching on the bias.}

The sample was mounted in a cryostat, and the temperature was varied between $280-330$~K using a thermoelectric cooler with a temperature step of $5$~K. The measurements were made in an argon atmosphere ($1000$~mbar).
\section{Results and Discussion}
\subsection{Ellipsometry}
\begin{figure}[!t]
\centering
\includegraphics[width=6cm]{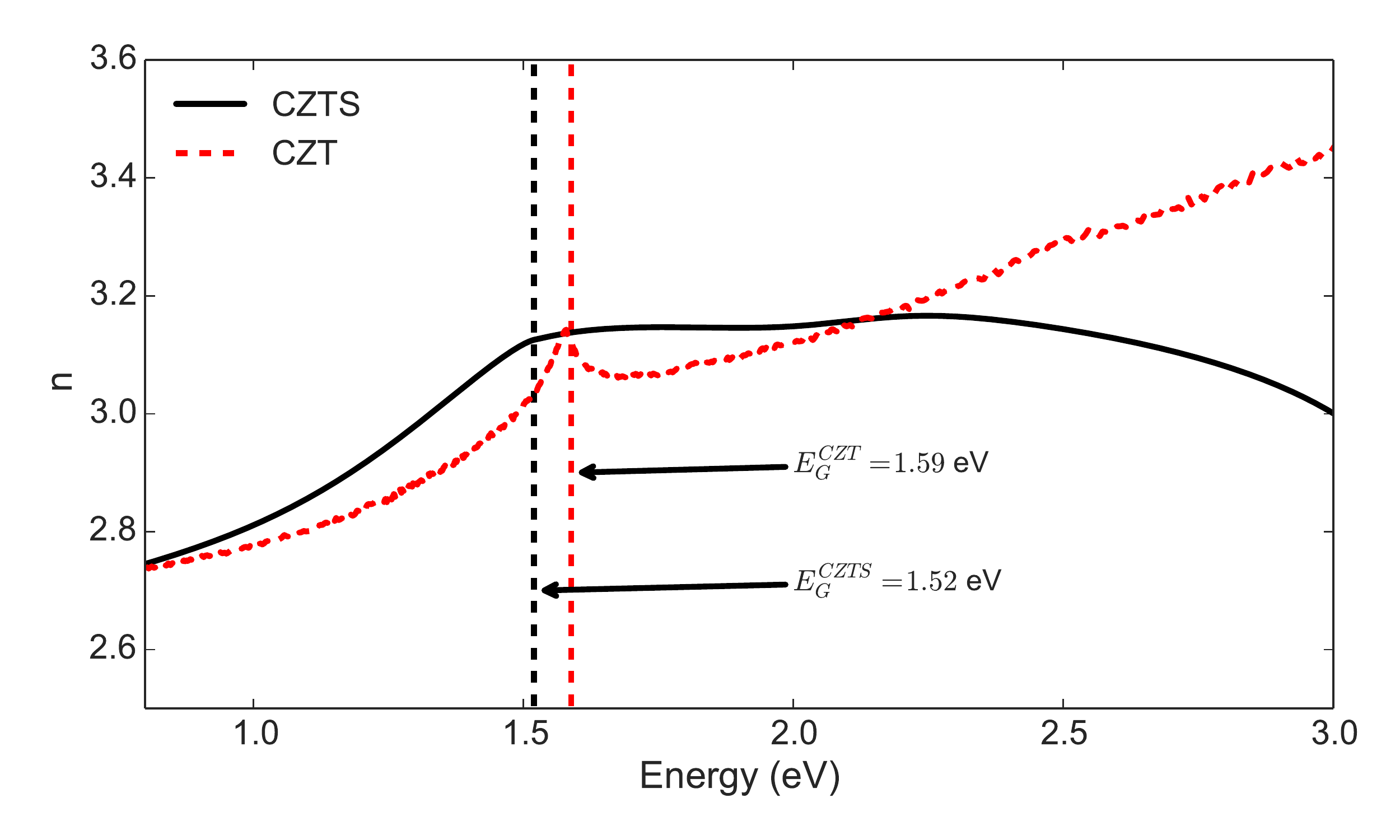}
\caption{\textcolor{black}{The optical refractive index of CZTS material (black solid line) and CZT  material (red dashed line). The determine energy bandgaps are marked by dashed lines.}}
\label{ellipsometry}
\end{figure}

\textcolor{black}{The spectroscopic ellipsometry measurements performed on the CZTS sample allows determination of the energy band.gap. We also measured the energy band-gap of the CZT material to determine the energy shift of the band-gap caused by the addition of Se. We fitted the experimental ellipsometry data by a theoretical model describing the CZTS/CZT materials. The CZTS/CZT materials were parametrized by a sum of Lorentz oscillators \cite{zazvorka2016,Rejhon2017}. The determined optical refractive index of CZTS bulk and CZT bulk is shown in figure \ref{ellipsometry}. The evaluated energy band-gap of CZTS is $E_{G}^{CZTS}=1.52$~eV, and for CZT it is $E_{G}^{CZT}=1.59$~eV.}

\subsection{Bias application}

\begin{figure*}[!t]
\centering
\includegraphics[width=12cm]{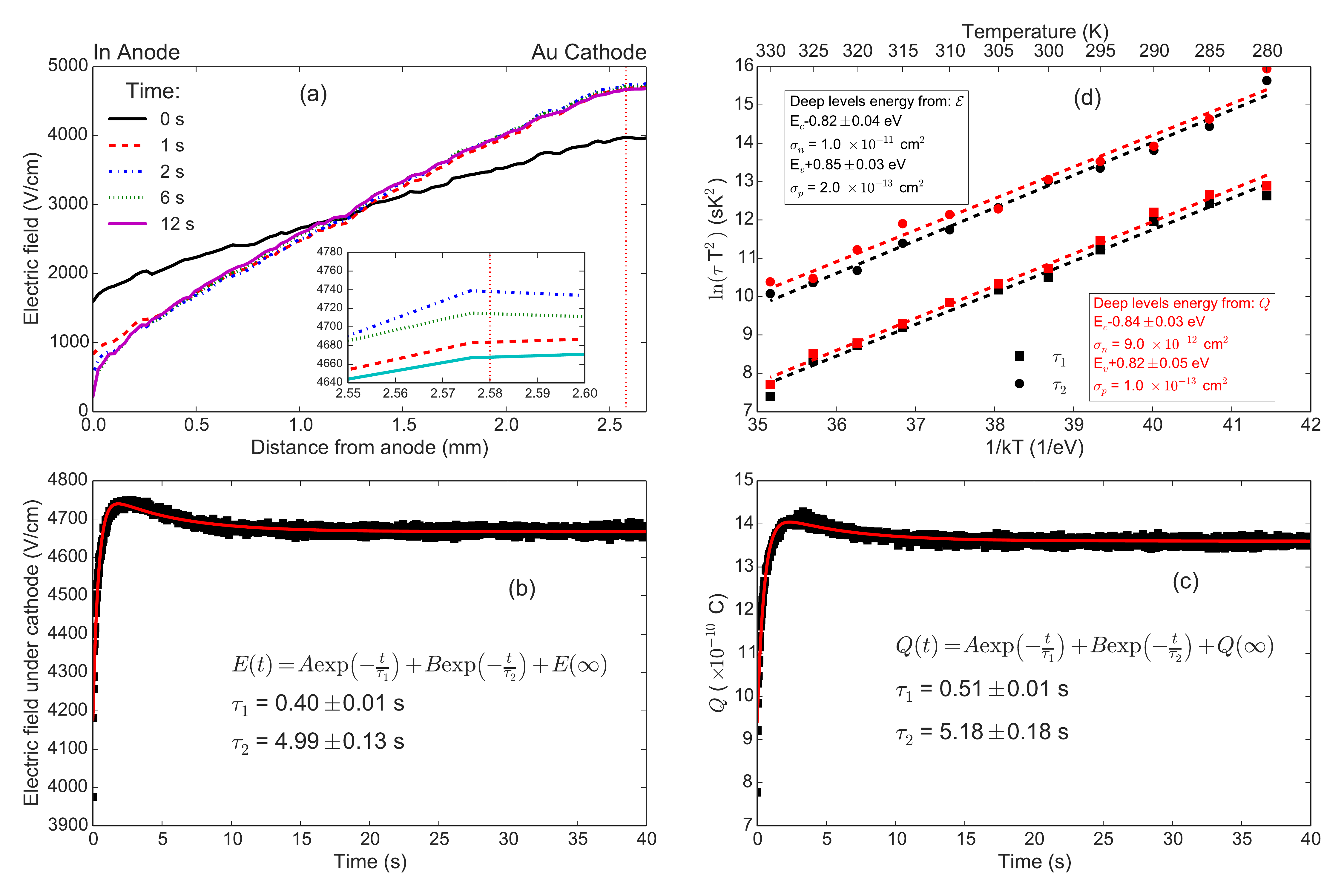}
\caption{The graph (a) shows the selected profile of electric field after bias application. The red dotted line shows the distance from the cathode. The time evolution of the electric field is plotted on the graph (b). The experimental data (black square) are fitted by a double-exponential function (red line) with time constants $\tau_{1}=0.40\pm0.01$~s and $\tau_{2}=4.99\pm0.13$~s. The space charge evolution in time is shown in graph (c). The data are shown as black squares, and the double-exponential fit with time constants is plotted as a red line. The graph (c) includes the determined time constants $\tau_{1}=0.51\pm0.01$~s and $\tau_{2}=5.18\pm0.18$~s. The Arrhenius diagram is plotted on the graph (d). The observed deep level energies are $E_{c}-0.82\pm0.04$~eV with electron capture cross-section $\sigma_{n}=1.0\times10^{-11}$~cm$^{2}$ and $E_{v}+0.85\pm0.03$~eV with hole capture cross-section $\sigma_{p}=2.0\times10^{-13}$~cm$^{2}$ from time constants describing the electric-field evolution. The deep levels determined from space charge evolution are $E_{c}-0.84\pm0.03$~eV with electron capture cross-section $\sigma_{n}=9.0\times10^{-12}$~cm$^{2}$ and $E_{v}+0.82\pm0.05$~eV with hole capture cross-section $\sigma_{p}=1.0\times10^{-13}$~cm$^{2}$.}
\label{bias}
\end{figure*}

The time evolution of the electric field in the sample after bias application was measured in the temperature range between $280-330$~K. The selected profiles of the electric field  are shown in figure \ref{bias}(a). The electric field in the steady state indicates the formation of a positive space charge in the sample.

The time evolution of the electric field under the cathode at $300$~K is plotted in figure \ref{bias}(b). The time evolution of the space charge $Q$ was computed for a better understanding of the electric field development by the formula \cite{Dedic2017}
\begin{equation}
Q=\varepsilon S(\mathcal{E}_{cathode}-\mathcal{E}_{anode}).
\label{eq_naboj}
\end{equation}
Here, $\varepsilon$ is the absolute permittivity, $S$ is the sample area, and $\mathcal{E}_{cathode,anode}$ is the electric field at the  cathode and the anode, respectively.

In our previous study of CZT \cite{Dedic2017}, we observed the undershoot in the evolution of the electric field under the cathode, but the space charge was described by a single-exponential function. It indicates a critically damped oscillation of the electric field caused by local space-charge fluctuations by one principal deep level. However, in the case of CZTS, the time evolution of the space charge (figure \ref{bias}(c)) exhibits a similar undershoot as the evolution of the electric field (figure \ref{bias}(b)). This effect cannot be explained by just one deep level. Therefore, we apply fitting of the experimental curves by a double-exponential function, which describes the influence of two dominant deep levels.

The activation energies of the levels $E_{i}$ were evaluated from the temperature dependence of the time constants $\tau_{i}$ describing the time evolution of the electric field under the cathode and space charge evolution by an Arrhenius equation
\begin{equation}
\ln(\tau_{i} T^{2})=\frac{E_{i}}{k_{B}T}+\ln\left(\frac{C}{\sigma_{i}}\right);\quad C=\frac{h^{3}}{16m^{\ast}_{e(h)}\pi k_{B}^{2}},
\end{equation}
where  $T$ is the absolute temperature, $k_{B}$ is the Boltzmann, $h$ is the Planck constant, $m^{\ast}_{e(h)}$ is the effective mass of electrons (holes) and $\sigma_{i}$  is the capture cross-section of the level.

We determined an activation energy of $0.82\pm0.04$~eV from the temperature dependence of $\tau_{1}$ describing the increase of the electric field under the cathode (black squares in figure \ref{bias}(d)). This process can be described by the thermal transition of electrons from the deep level to the conduction band resulting in the observed increase of the positive space charge. Therefore, we assigned the activation energy to the deep level $E_{1}=E_{c}-0.82\pm0.04$~eV with electron capture cross section $\sigma_{n}=1.0\times10^{-11}$~cm$^{2}$.

The time constant $\tau_{2}$ describes the decrease of the electric field under the cathode, which correlates with the electron transition from the valence band to the deep level. The data are plotted as black circles in figure \ref{bias}(d). The determined deep level is $E_{2}=E_{v}+0.85\pm0.03$~eV with hole capture cross section $\sigma_{p}=2.0\times10^{-13}$~cm$^{2}$.

The Arrhenius analysis was performed on time constants describing the space-charge evolution. The time constant $\tau_{1}$ related to the increase of positive space charge leads to the deep level $E_{1}=E_{c}-0.84\pm0.03$~eV with electron capture cross section $\sigma_{n}=9.0\times10^{-12}$~cm$^{2}$ (red squares in figure \ref{bias}(d)). The analysis of the time constant $\tau_{2}$ (red circles in figure \ref{bias}(d)) provides the deep level $E_{2}=E_{v}+0.82\pm0.05$~eV with hole capture cross section $\sigma_{p}=1.0\times10^{-13}$~cm$^{2}$ due to the decreasing of the positive space charge. We conclude that both methods of evaluation (electric field and total space charge) lead to the same results within the expected experimental error.

\begin{figure*}[!t]
\centering
\includegraphics[width=12cm]{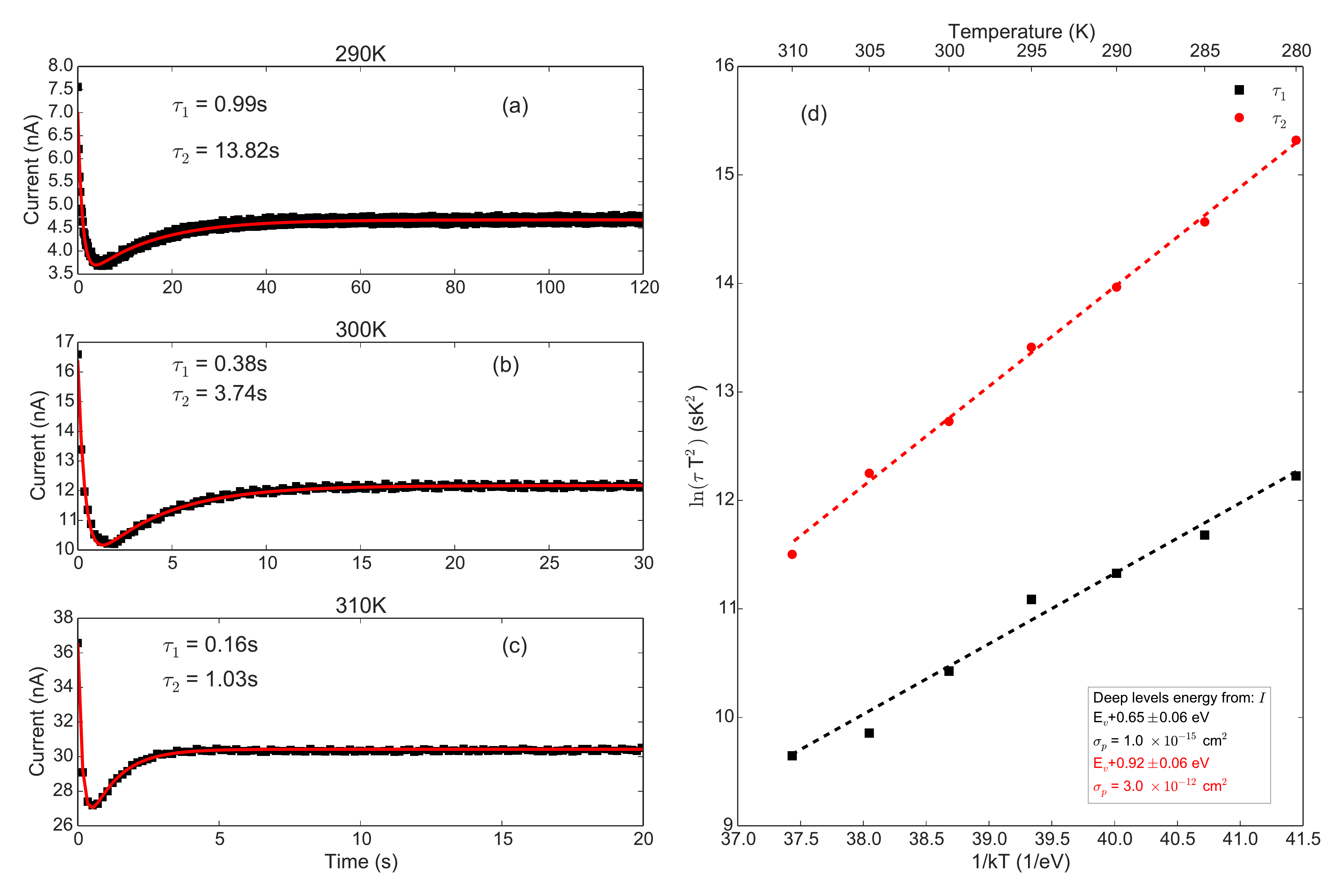}
\caption{\textcolor{black}{The graphs (a-c) show the time evolution of the measured electric current after switching on the bias at three different temperatures $290$ (a), $300$ (b) and $310$~K (c). The Arrhenius diagram is plotted on graph \ref{current}(d). The observed deep-level energies are $E_{v}-0.65\pm0.06$~eV with hole capture cross-section $\sigma_{p}=1.0\times10^{-15}$~cm$^{2}$ and $E_{v}+0.92\pm0.06$~eV with hole capture cross-section $\sigma_{p}=3.0\times10^{-12}$~cm$^{2}$.}}
\label{current}
\end{figure*}

\textcolor{black}{Figures \ref{current}(a-c) show the time evolutions of the measured electric current after switching on the bias at three different temperatures $290$, $300$ and $310$~K. The current decreases within the first two seconds with time constant $\tau_{1}=0.38$~sec at $300$~K. It is the same time constant with which the electric field below the cathode increases (figure \ref{bias}(b)). The buildup of positive space charge is thus accompanied with a decrease of the electric current. The Arrhenius analysis the time constant $\tau_{1}$ (figure \ref{current}(d)) gives an energy of $0.65$~eV and carrier capture cross-section of $1\times10^{-15}$~cm$^{2}$. This energy is complementary to the electron trap $E_{1}=E_{c}-0.83$~eV. The sum $0.83+0.65$~eV agrees within the experimental error with the band-gap of the material at $1.52$~eV as measured by ellipsometry. This energy is therefore connected with thermally activated emission of holes from the valence band to the level causing that the absolute value of the electric current immediately after switching on the bias increases with temperature (from $7.5$~nA at $290$~K to $38$~nA at $310$~K).}

\textcolor{black}{In summary, within the first two seconds after application of the bias, the positive space charge is built  up by electron transitions from this level to conduction band. At the same time free holes are trapped at the level decreasing the electric current, which therefore seems to be composed mainly from holes. After the initial decrease the electric current starts to increase with time constant $\tau_{2}=3.74$~sec at $300$~K. An Arrhenius analysis of this time constant gives an activation energy of $0.92$~eV. With a similar time activation energy and time constant, the  electric field below the cathode decreases (positive space charge decreases - figure \ref{bias}(c)). Therefore, we conclude that the observed increase of the electric current is caused by the emission of holes at the level $E_{2}=E_{v}+0.83$~eV. Corresponding transitions of electrons from the valence band to the level cause the decrease of the positive space charge.}

\textcolor{black}{Additional analysis of time and temperature evolution of the electric current fully confirmed the conclusions obtained from evaluation of the electric-field dependences. Furthermore, the capture cross-section for holes at the electron trap $E_{1}=E_{c}-0.83$~eV could be evaluated from the data.}
\subsection{Infrared spectral scanning}

\begin{figure}[!t]
\centering
\includegraphics[width=6cm]{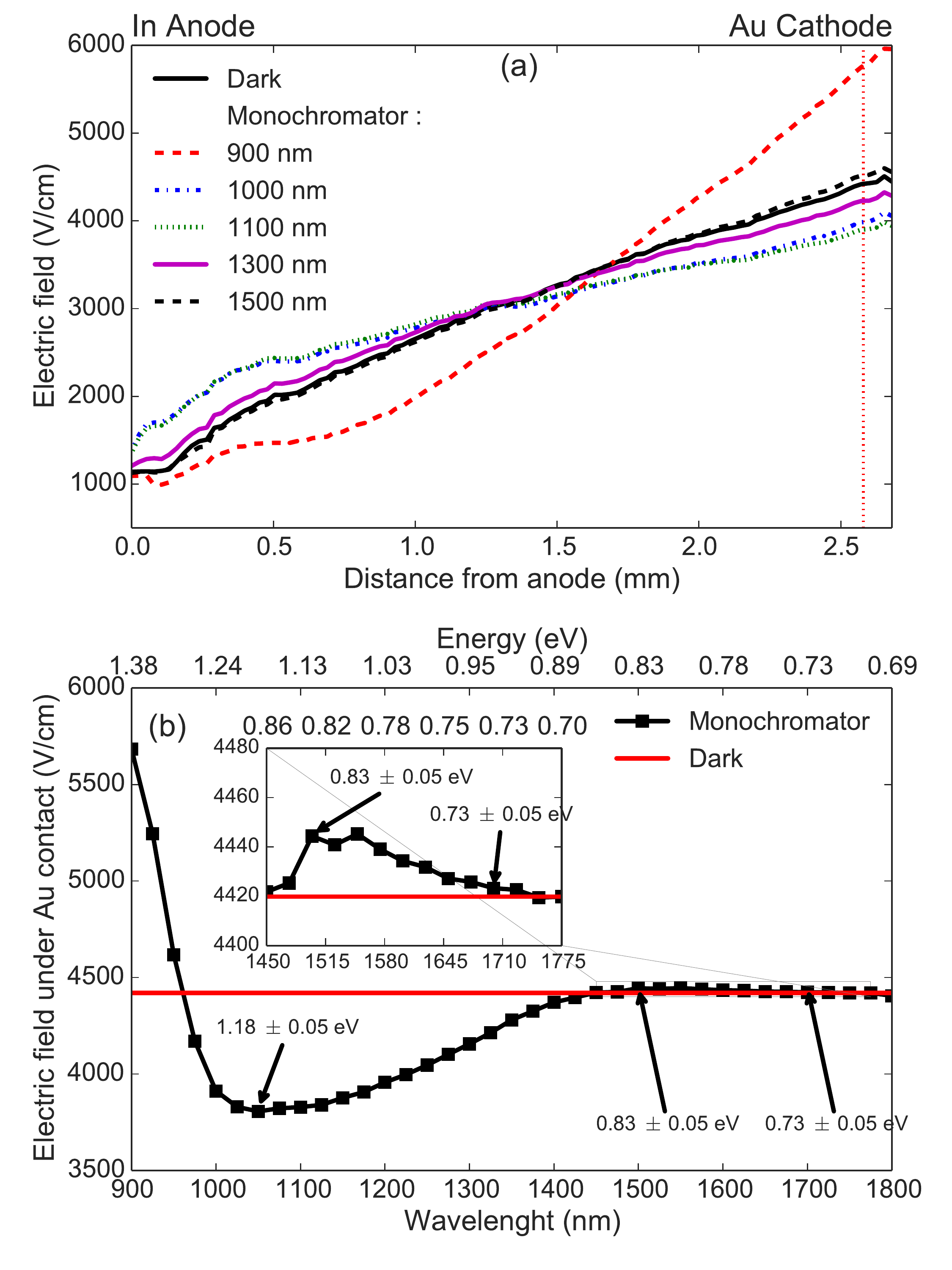}
\caption{The profiles of the electric field with and without illumination from the source of the monochromator light (a). The applied voltage was $800$~V on the In contact and the temperature was $300$~K. The red dotted line demonstrates the distance from the cathode where the electric field was cut and plotted showing the dependence on illumination wavelength on the bottom figure (b). The inset graph shows the area in the range of $1450-1775$~nm. The arrows mark the change of the electric field under the cathode induced by the optical transition. The determined deep levels are $0.73$, $0.83$ and $1.18$~eV with a $0.05$~eV error.}
\label{sken}
\end{figure}

The dependence of the selected electric-field profiles on the NIR illumination from the monochromator is shown in figure \ref{sken}(a). The electric-field profile without illumination linearly increases from the anode to the cathode, which indicates a positive space charge in the sample volume. The below band-gap NIR light from the monochromator penetrates through the whole sample volume. The penetrated light can affect the occupation of deep levels, which leads to a change of the space-charge distribution and modification of the electric-field profile.

The dependence of the electric field under the cathode ($0.1$~mm from cathode) is shown in figure \ref{sken}(b) for a step of $25$~nm. The illumination at $1725$~nm ($0.73$~eV) causes a slight increase of the electric field under the cathode due to transitions of electrons from the deep level to the conduction band $E_{3}=E_{c}-0.73$~eV. This transition causes an increase of the positive space charge. It leads to the increase of the electric field under the cathode. Other optically induced electron transitions affecting the electric field start at $1500$~nm ($0.83$~eV). The decrease of the electric field under the cathode indicates a decrease of the positive space charge in the sample. It corresponds to transitions of electrons from the valence band to the deep level $E_{2}=E_{v}+0.83$~eV. The electric field under the cathode starts to increase at $1050$~nm ($1.18$~eV) due to electron transitions from the deep level at $E_{4}=E_{c}-1.18$~eV to the conduction band. This transfer forms the positive space charge in the sample.

\subsection{Time and temperature evolution of the electric field after switching off the LED at 940 nm}

We monitored the time evolution of the electric field after switching off the LED at the wavelength of 940~nm. The illumination by $940$-nm light causes an increase of the positive space charge in the sample. The electric field returns to the dark condition (steady state) after switching off the 940-nm LED. It means that the generated positive space charge is reduced due to the thermally activated electron transitions from the valence band to the deep level or due to the non-thermally activated electron capture from the conduction band to the deep level.

\begin{figure*}[!t]
\centering
\includegraphics[width=12cm]{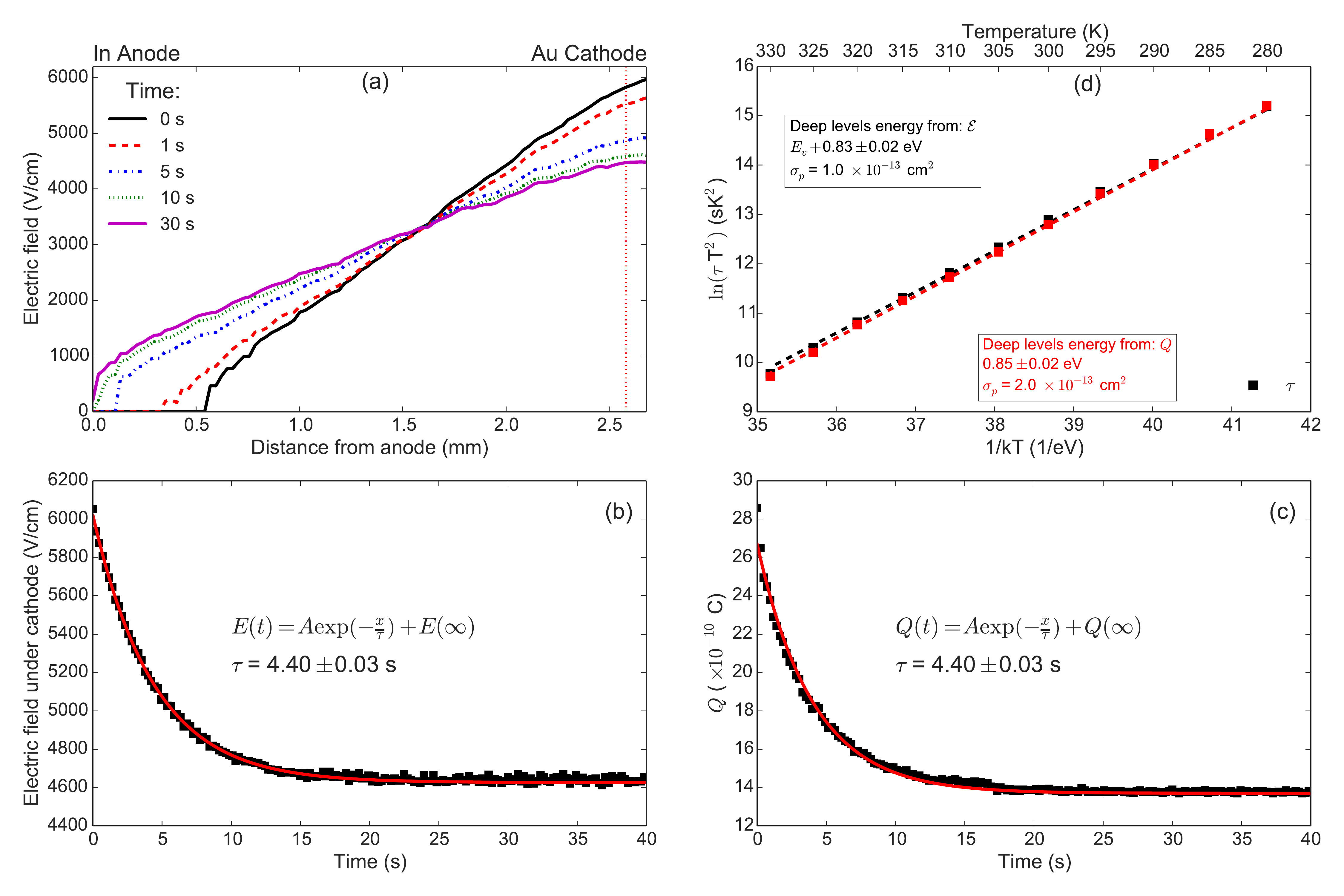}
\caption{The graph (a) shows the selected profiles of the electric field after switching off the LED at $940$~nm at a temperature of $300$~K and bias of $800$~V on the In contact. The red dotted line shows the distance from cathode, and the time evolution of the electric field is plotted in graph (b). This graph shows a mono-exponential decay of the electric field under the cathode (black square) with the mono-exponential fit (red line). The graph is supplemented by the determined time constant $\tau=4.40\pm0.03$~s. The computed space charge is plotted on the graph (c) as black squares with the single exponential fit (red line). The Arrhenius diagram is plotted on the graph (d). The observed deep level energy is $E_{v}+0.83\pm0.02$~eV ($E_{v}+0.85\pm0.02$~eV) with hole capture cross-section of $\sigma_{p}=1.0\times10^{-13}$~cm$^{2}$ ($\sigma_{p}=2.0\times10^{-13}$~cm$^{2}$) from the evolution of the electric field under the cathode (space charge evolution).}
\label{led940}
\end{figure*}

The selected electric-field profiles at $0$, $1$, $5$, $10$ and $30$~seconds after switching off the LED operating at $940$~nm are shown in figure \ref{led940}(a). The electric-field profile at $0$~second indicates a dead layer under the In anode due to the generated positive space charge.

The corresponding time evolution of the electric field under the cathode is depicted in figure \ref{led940}(b), and the computed space charge according to equation (\ref{eq_naboj}) is plotted in figure \ref{led940}(c). Both time dependences exhibit a single exponential decay.

The activation energy of thermally activated transitions is identified using an Arrhenius plot (figure \ref{led940}(d)). The transition from the valence band to the deep level $E_{2}=E_{v}+(0.83\pm0.02)$~eV with a capture cross-section of $\sigma_{p}=1\times10^{-13}$~cm$^{2}$ is evaluated from the time constant of the corresponding electric-field evolution. The temperature dependence of the time constant describing the space-charge evolution leads to the deep level $E_{2}=E_{v}+(0.85\pm0.02)$~eV with a capture cross-section of $\sigma_{p}=2\times10^{-13}$~cm$^{2}$.

The main deep level in the CZTS sample responsible for the optically induced polarization is at $E_{2}=E_{v}+0.83$~eV. The incident light at $940$~nm caused the electron transitions from this level to the conduction band. It increased the positive space charge, which leads to the polarization of the detector. After switching off the illumination at $940$~nm, the electron returns to this deep level from the valence band. It causes the decrease of the induced positive space charge. We conclude that both the energy and capture cross-section of this deep level are  the same as those evaluated after bias application.

\subsection{Comparison with CdZnTe}

\begin{figure*}[!t]
\centering
\includegraphics[width=12cm]{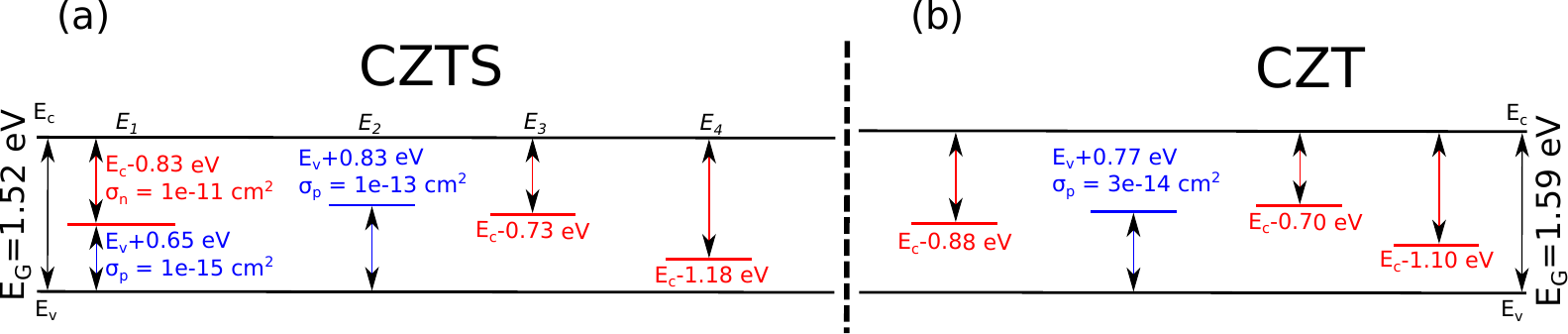}
\caption{Schematic of the observed energy levels with their capture cross-section determined by temperature measurements and by an infrared spectral scanning method (graph (a)). The red color represents the electron traps, and blue represents hole traps. We supplemented the scheme of the deep levels in CZT material observed in references \cite{rejhonJPD,Franc2015,Zazvorka2014} (graph (b)).}
\label{porovnani}
\end{figure*}

The observed deep levels in CZTS material are shown in the energy scheme (figure \ref{porovnani}(a)). The CZT deep levels evaluated by the same approach \cite{rejhonJPD,Franc2015,Zazvorka2014} are presented in figure \ref{porovnani}(b). The deep level at $E_{4}=E_{c}-1.18$~eV was evaluated from spectral infrared scanning in the CZTS material. This deep level is responsible for strong optically induced polarization. The CZT material demonstrates the deep level at $E_{c}-1.10$~eV determined by the same method \cite{Franc2015,Zazvorka2014}. Both start to form positive space charge in the material, so we assume that the levels have the same origin. Kim \cite{Kim2013} assigned this deep level to dislocations induced by Te inclusions/precipitates. Castaldini \cite{Castaldini1998} attributed this energy level at $1.1$~eV below the conduction level to a positively charged tellurium vacancy. We assume that the observed energy shift is caused by the addition of Se content.

\textcolor{black}{It is apparent that the activation energies of electrons from electron traps  $E_{3}$ and $E_{4}$ to the conduction band are higher in CZTS  by $30-80$~eV when compared to CZT. The activation energy of electrons to the valence band of the hole trap $E_{2}$ is also higher by approximately $50$~meV in case of CZTS. At the same time the energy band of CZTS is smaller by $70$~meV compared to CZT (figure \ref{ellipsometry}). It means that the observed shifts of energies in CZTS when compared to CZT cannot be simply explained by the change of the band-gap after introduction of Se in the lattice. While the change of energies of the principal deep levels may seem relatively small, it can significantly influence occupations of these levels if they are close to the Fermi energy, and in  this way affect the stability of the Fermi level along the ingot as well as trapping and recombination of free carriers. A detailed study of the impact of these energy-level shifts will be a focus of our future research.}

The deep level $E_{2}=E_{v}+0.83$~eV in CZTS has probably the same origin as the level $E_{v}+0.77$~eV in CZT \cite{rejhonJPD}. This level was assigned to the second ionized state of cadmium vacancy by Castaldini \cite{Castaldini1998}. It is responsible for both a thermally and an optically induced decrease of positive space charge after application of the bias and infrared light. The observed energy shift of the level towards the conduction band in CZTS compared to CZT is in agreement with theoretical predictions of Varley \cite{Varley2017}. They used first-principles calculations based on screened hybrid functionals and found that cation vacancies shifted towards the conduction band  when Se replaces Te in the lattice. The observed energy shift therefore also strongly indicates that the studied defect is related to Cd vacancies. 

The deep level $E_{1}=E_{c}-0.83$~eV observed from temporal and temperature analysis is responsible for setting the electric field after the application of the bias in CZTS. Also, it is responsible for the positive space charge at steady state.

The deep level $E_{3}=E_{c}-0.73$~eV in CZTS could have the same origin as the deep level $E_{c}-0.70$~eV in CZT material. These levels exhibit weak optically induced electron transitions from these deep levels to the conduction band.
\section{Conclusions}
We analyzed the deep levels in CZTS material responsible for the dynamics of the electric-field profile and electric current by a complex approach involving the optical and temporal and temperature measurements of the electrical field using the electro-optic Pockels effect. We identified three electron deep levels at $E_{1}=E_{c}-0.83$~eV, $E_{3}=E_{c}-0.73$~eV and $E_{4}=E_{c}-1.18$~eV and one hole deep level at $E_{2}=E_{v}+0.83$~eV. The electron deep level $E_{1}=E_{c}-0.83$~eV is responsible for the positive space charge in the sample. We show that the sample can be depolarized using the hole trap at $E_{2}=E_{v}+0.83$~eV. The electron deep level $E_{4}=E_{c}-1.18$~eV is responsible for the strong optically induced polarization of the sample due to the trapping of photo-generated holes, which form a positive space charge.

The deep levels in CZTS are compared to those measured in CZT by the same methods. We observed the theoretically predicted an energy shift of the second ionization level of the cadmium vacancy towards the conduction band in CZTS relative to CZT.

\section*{Acknowledgment}

This paper was financially supported by the  Grant Agency of Czech Republic (GA\v{C}R), project 102-18-06818S and student project SVV–2018–260445.

\ifCLASSOPTIONcaptionsoff
  \newpage
\fi



\bibliographystyle{IEEEtran}
%



%

\end{document}